\pdfoutput=1
\documentclass[11pt]{article}

\usepackage[]{acl}

\usepackage{times}
\usepackage{latexsym}

\usepackage{graphicx}
\usepackage{multirow}
\usepackage{amsmath,bm}
\usepackage{amssymb}
\usepackage{xcolor}
\usepackage{comment}
\usepackage{algorithm}
\usepackage{algpseudocode}
\usepackage{hyperref}

\usepackage{tabularx}

\usepackage{stmaryrd}

\usepackage{cleveref}
\usepackage[normalem]{ulem}
\usepackage{soul}
\definecolor{mypink1}{rgb}{0.858, 0.188, 0.478}
\definecolor{mygreen}{rgb}{0, 1, 0}
\definecolor{mymaroon}{rgb}{0.5, 0, 0}
\definecolor{mypurple}{rgb}{0.5, 0, 0.5}
\definecolor{myauburn}{rgb}{0.43, 0.21, 0.1}

\newcommand{\hqcmt}[1]{{\color{mymaroon}{\emph{#1}}}}

\renewcommand{\hqcmt}[1]{}

\usepackage[T1]{fontenc}
\usepackage[utf8]{inputenc}

\usepackage{microtype}

\usepackage{inconsolata}

\title{Zero-Shot End-to-End Spoken Language Understanding\\
via Cross-Modal Selective Self-Training}

 \author{
Jianfeng He{$^{1,2}$}\thanks{~The work was done during an AWS AI Labs internship.}~, Julian Salazar{$^1$}, Kaisheng Yao{$^1$}, Haoqi Li{$^1$}, Jason Cai{$^1$}
\\ 
   {$^1$} AWS AI Labs\\
 {$^2$} Virginia Tech\\
 jianfenghe@vt.edu, cjinglun@amazon.com
 }

\begin{document}
\maketitle
\begin{abstract}
End-to-end (E2E) spoken language understanding (SLU) is constrained by the cost of collecting speech-semantics pairs, especially when label domains change. Hence, we explore \textit{zero-shot} E2E SLU, which learns E2E SLU without speech-semantics pairs, instead using only speech-text and text-semantics pairs. 
Previous work achieved zero-shot by pseudolabeling all speech-text transcripts with a natural language understanding (NLU) model learned on text-semantics corpora. 
However, this method requires the domains of speech-text and text-semantics to match, which often mismatch due to separate collections. Furthermore, using the entire collected speech-text corpus from any domains leads to \textit{imbalance} and \textit{noise} issues. To address these, we propose \textit{cross-modal selective self-training} (CMSST). CMSST tackles imbalance by clustering in a joint space of the three modalities (speech, text, and semantics) and handles label noise with a selection network. We also introduce two benchmarks for zero-shot E2E SLU, covering matched and found speech (mismatched) settings. Experiments show that CMSST improves performance in both two settings, with significantly reduced sample sizes and training time. Our code and data are released in \url{https://github.com/amazon-science/zero-shot-E2E-slu}.

\end{abstract}

\begin{figure*}[!htbp]
\centering
\includegraphics[width=\textwidth]{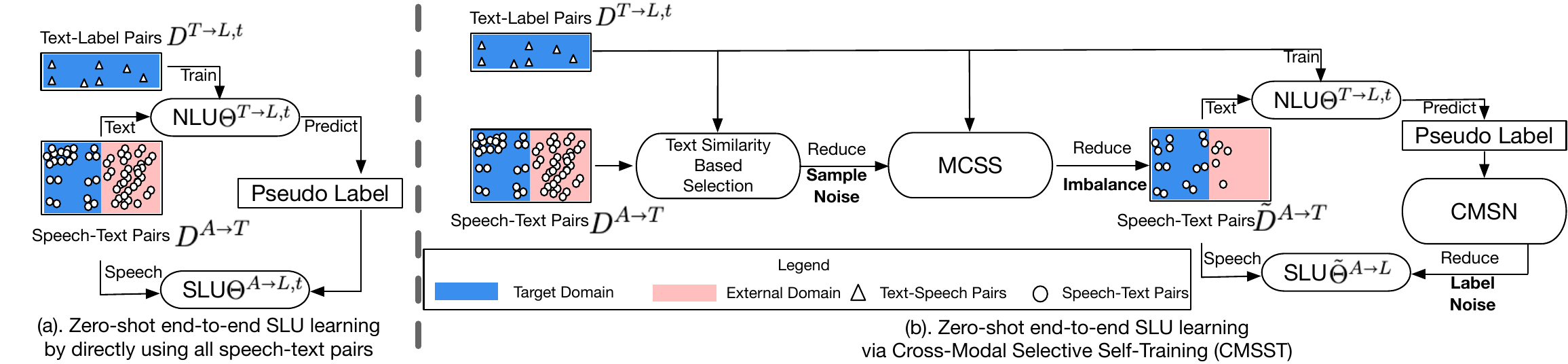}
\caption{\textbf{(a)}. Diagram of using all speech-text pairs, detailed in Sec.~\ref{sec:intro}. The legend in (b) is also applicable to (a). 
\textbf{(b)}. Diagram of the CMSST framework (described in Sec.~\ref{sec:framework}). 
Speech and text pairs in $D^{A \shortrightarrow T}$ are selected by first using a text-similarity-based selection method and then a Multi-view Clustering-based Sample Selection (MCSS) algorithm. The SLU model $\tilde{\Theta}^{A \shortrightarrow L}$ is trained on the resulting speech-text pairs $\tilde{D}^{A \shortrightarrow T}$, with pseudolabels from an NLU model $\Theta^{T\shortrightarrow L, t}$. This NLU model is trained from target domain text-to-semantics pairs $D^{T\shortrightarrow L, t}$. To deal with label noise from the NLU model, CMSST uses a Cross-Modal SelectiveNet (CMSN) to train our SLU model $\tilde{\Theta}^{A \shortrightarrow L}$.
}
\label{fig:overview}
\end{figure*}

\section{Introduction}
\label{sec:intro}

End-to-end (E2E) spoken language understanding (SLU) models train on speech-semantics pairs, inferring  semantics directly from acoustic features~\cite{serdyuk2018towards} and leveraging non-lexical information like stress and intonation. In contrast, pipelined SLU models~\cite{tur2011spoken} operate on speech-transcribed text, omitting the acoustic information.
In all, E2E SLU has gained significant research attention. 
However, training E2E SLU models faces a significant challenge in collecting numerous speech-semantics pairs~\cite{hsu2021hubert}. 
This challenge is two-fold: the scarcity of public speech-semantics pairs due to annotation costs and the need to relabel speeches when the labeling schema evolves, e.g., functionality expansion~\citep{goyal2018fast}.
While speech-semantics pairs are scarce and expensive to annotate, there is a growing availability of speech-text pairs used in automatic speech recognition (ASR) and text-semantics pairs used in natural language understanding (NLU) \cite{galvez2021people,fitzgerald2022massive}. Thus, we define \textit{zero-shot} E2E SLU, which learns an E2E SLU model by speech-text and text-semantics pairs \textit{without ground-truth speech-semantics pairs} (hence zero-shot, a more detailed explanation is given in Sec.~\ref{sec:why_zero_shot}). 

Two works have explored zero-shot E2E SLU. \citet{pasad-etal-2022-use} trained an NLU model by text-semantics pairs and used it to predict pseudolabels for the text of \textit{all} speech-text pairs, similar to Figure~\ref{fig:overview}(a). They then trained an E2E SLU model using the speech audio from the speech-text pairs, paired with the predicted pseudolabels. In another way, \citet{mdhaffar2022e2e} mapped the text of \textit{all} text-semantics pairs to speech embeddings, creating ``pseudospeech''-semantics pairs.

However, both works assume matched domains for text-semantics and speech-text pairs, with data collected from the same scenario. In practice, however, these pairs are often separately collected, leading to potential domain mismatches. In such cases, directly using all speech-text and text-semantics pairs for zero-shot E2E SLU leads to two types of issues as below.
\\
\textbf{Noise.} \textit{Sample noise} comes from speech-text pairs whose transcripts (texts) are out-of-domain (OOD) for the NLU task. 
Passing all transcripts through NLU inference leads to inaccurate pseudolabels on the OOD data, impacting SLU learning. This exacerbates \textit{label noise}, which refers to incorrect NLU model predictions that are then (wrongly) treated as pseudolabels; this issue is inherent to self-training and also impacts performance~\cite{du2020self}.
\\
\textbf{Imbalance.} Since the text-semantics and speech-text pairs are separately collected, even after removing OOD speech-text pairs, the remaining text in speech-text pairs may be heavily imbalanced within the NLU domain, e.g., one semantics dominates all others. Besides, imbalanced speech, e.g., having only female voices, can bias E2E SLU learning. 
Though a model may succeed despite the imbalance, this can waste training resources that could have been used on representative speech-text pairs.

For these issues, \citet{pasad-etal-2022-use} and \citet{mdhaffar2022e2e} ignore sample noise and imbalance by selecting speech-text pairs that are matched and balanced; however, in practice, it is hard to gain such well-matched and well-balanced speech-text corpus. Furthermore, neither work is selective with pseudodata, which in \citet{pasad-etal-2022-use} led to degradation when more external speech-text was added, due to label noise. Instead, with \textit{selection} as a unifying perspective, we make the following contributions:
\\
\textbf{(i). Zero-shot E2E SLU benchmarks for both matched and found speech.} For the matched domain setting, we define \textbf{VoxPopuli2SLUE}, combining text-semantics pairs of SLUE's NER-annotated subset \citep{shon2022slue} of VoxPopuli~\citep{wang2021voxpopuli} with speech-text pairs from VoxPopuli, similar to \citet{pasad-etal-2022-use}. Then, for the found (mismatched) speech setting, we define \textbf{MiniPS2SLURP}, combining the home-assistant text-semantics pairs of SLURP \citep{bastianelli2020slurp} with speech-text pairs from the general-domain People's Speech corpus~\citep{galvez2021people}. 
\\
\textbf{(ii). Selection via cross-modal clustering and selective networks to tackle imbalance and noise in self-training.}  To tackle sample noise, we first exclude OOD speech-text pairs using text similarity. Then, for the imbalance, we propose \textit{multi-view clustering-based sample selection (MCSS)} to resample speech-text pairs to improve diversity over three views (speech, text and latent semantics). For label noise, we propose a \textit{cross-modal SelectiveNet (CMSN)}, which selectively trusts pseudolabels based on the ease of learning common representations between the NLU and SLU encoders. All together, we refer to our proposed framework as \textbf{cross-modal selective self-training (CMSST)}, summarized in Figure~\ref{fig:overview}(b).
\\
\textbf{(iii). Comprehensive experiments on zero-shot E2E SLU.} We compare the baselines with our CMSST on the new benchmarks. CMSST achieves better results with significantly less data. Ablations show that clustering and selective learning both contribute; Entity F1 improves 1.2 points on VoxPopuli2SLUE with MCSS and 1.5 points on MiniPS2SLURP with CMSN.

\section{Related Work}

\textbf{Speech to semantics.} Although not fully zero-shot, works in semi-supervised E2E SLU have also considered the mismatch problem. \citet{Rao2020s2s} train NLU and ASR systems independently, saving their task-specific SLU data for a final joint training stage.
Others tackle the data sparsity or mismatch issues using text-to-speech (TTS) to synthesize spoken counterparts to NLU examples \citep{Lugosch2019ttsnlu,lu2023improving}. Pretraining on off-the-shelf (found) speech-only data \citep{lugosch2019speech}, text-only data \citep{huang2020leveraging}, or both \citep{chung2020splat,thomas2022towards} 
have improved SLU systems beyond their core speech-semantics training data, 
usually via an alignment objective or joint network. Finally, \citet{rongali2021exploring} considered a different notion of ``zero-shot'' E2E SLU, which we view more aptly as text-only SLU adaptation; their setting involves an initial E2E SLU model, trained on speech-semantics pairs, having its label set expanded with text-only data.
\\
\noindent \textbf{Self-training.} This method \citep{scudder1965probability, yarowsky1995unsupervised} further trains a model on unlabeled inputs that are labeled by the same model, as a form of semi-supervised learning. It has experienced a recent revival in both ASR \cite{kim2023asbert} and NLU \cite{le2023reducing}, giving improvements atop strong supervised and self-supervised models, for which effective sample filters and label confidence models were key. Recently, \citet{pasad-etal-2022-use} performed self-training in the zero-shot E2E NER case; however, since they work in the matched case they do not address these issues of imbalance and noise.
\\
\textbf{Multi-view clustering.} Multiple views of the data can improve clustering by integrating extensive information~\cite{kumar2011co,wang2022highly,fang2023dbo,huang2023fast}. We propose using the modalities in speech-text pairs (speech, text, and latent semantics) as bases to build a joint space, where we apply clusters to enable balanced selection. We apply simple heuristics atop the clusters, and leave stronger algorithms, e.g., \citet{trosten2021reconsidering} to future work.
\\
\textbf{Selective learning.} Selective learning aims at designing models that are robust in the presence of mislabeled datasets~\cite{ziyin2020learning}. 
It is often achieved by a selective function~\cite{geifman2019selectivenet}.
Selective learning has been recently applied in a variety of applications~\cite{chen2023selective,kuhne2022defending,chen2023interpretable}. But less so in NLP applications~\cite{xin2021art} and little in cross-modal areas. 

\begin{table}[!tbh]
    \centering
    \scriptsize
    \begin{tabular}{llrr} 
    \hline
    &  & \textbf{MiniPS2} & \textbf{VoxPopuli2}  \\
    \textbf{Data} & \textbf{Annotation}  & \textbf{SLURP} & \textbf{SLUE}  \\
    \hline
    \multirow{2}{*}{$D^{A \shortrightarrow L, t}$} & Speech-to-semantics pairs  &  \multirow{2}{*}{22,782} &  \multirow{2}{*}{2,250}  \\
     & in target domain $t$ &   &    \\
    \hline
    \multirow{2}{*}{$D^{T \shortrightarrow L, t}$} & Text-to-semantics pairs   &  \multirow{2}{*}{22,783} & \multirow{2}{*}{2,250}   \\
     & in target domain $t$  &   &   \\
    \hline
    \multirow{2}{*}{$D^{A \shortrightarrow T, t} $} & Speech-to-text pairs & \multirow{2}{*}{22,782} & \multirow{2}{*}{2,250}  \\ 
     & in target domain $t$ &  &  \\ 
     \hline
    \multirow{2}{*}{$ D^{A \shortrightarrow T, \epsilon}$} & Speech-to-text pairs  & \multirow{2}{*}{32,255} & \multirow{2}{*}{182,466} \\
     & in external domains $\epsilon$  &  &  \\
     \hline
    \multirow{2}{*}{$D^{A \shortrightarrow T}$} & Union of $D^{A \shortrightarrow T, t}\ $   & \multirow{2}{*}{55,037} & \multirow{2}{*}{184,716}  \\ 
     & and $D^{A \shortrightarrow T, \epsilon}$ &  &   \\ 
    \hline
    \multirow{2}{*}{Test} & Test speech-to-semantics  & \multirow{2}{*}{13,078} & \multirow{2}{*}{877} \\
    & pairs in target domain $t$ &  &  \\
    \hline 
    \end{tabular}
    \caption{Data annotations and sample sizes in our datasets.  $D^{A \shortrightarrow L, t}$ is used for training a target SLU model $\Theta^{A \shortrightarrow L, t}$. $D^{T \shortrightarrow L, t}$ and $D^{A \shortrightarrow T}$ are used for training our E2E SLU model $\tilde{\Theta}^{A \shortrightarrow L}$. } 
    \label{tab:dataset_stats}
\end{table}

\section{Benchmarks for Zero-Shot E2E SLU}
\label{sec:prbolem_set}
We define a traditional SLU model as $\Theta^{A \shortrightarrow L, t}$, that is trained on data $D^{A \shortrightarrow L, t}$ with pairs of speech \textbf{audio $A$} and semantic \textbf{labels $L$}. These samples are in a target domain $t$. Besides, we will use superscript $T \shortrightarrow L$ to denote \textbf{text $T$} to semantic labels, and $A \shortrightarrow T$ to denote 
speech audio to text.

In our zero-shot setting, instead of having a speech-to-semantics dataset $D^{A \shortrightarrow L, t} $, we have a text-to-semantics pair set $D^{T \shortrightarrow L, t}$ in the target domain, and an external speech-to-text pair set $D^{A \shortrightarrow T}$.
Unlike~\citet{pasad-etal-2022-use} or \citet{mdhaffar2022e2e}, the provided speech-to-text data $D^{A \shortrightarrow T}$ may be independently collected and have sample pairs from an external domain. We divide $D^{A \shortrightarrow T}$ into two disjoint subsets, with samples either in the \textbf{target domain $t$} or being \textbf{external domain $\epsilon$}:
\begin{equation}
D^{A \shortrightarrow T} = D^{A \shortrightarrow T, t} \cup D^{A \shortrightarrow T, \epsilon}.
\end{equation}
A domain denotes data collection scenarios.
The $\epsilon$ can be \textit{matched} or \textit{mismatched} to the $t$ domain.

Given $D^{T \shortrightarrow L, t}$ and $D^{A \shortrightarrow T}$, we aim to learn an E2E SLU model $\tilde{\Theta}^{A \shortrightarrow L}$ that performs close to ${\Theta}^{A \shortrightarrow L, t}$.
This is zero-shot, as training our $\tilde{\Theta}^{A \shortrightarrow L}$ uses no speech-semantics pairs $D^{A \shortrightarrow L, t}$. We created the below two datasets to study this problem:

\textbf{Matched Speech: VoxPopuli2SLUE.}
We use \textit{SLUE-VoxPopuli}~\cite{shon2022slue} as the target domain text-to-semantics data $D^{T \shortrightarrow L, t} $. The external speech-to-text data $D^{A \shortrightarrow T}$ is from \textit{VoxPopuli}~\cite{wang2021voxpopuli}. We denote this dataset as VoxPopuli2SLUE. Its domain is matched, because SLUE-VoxPopuli and VoxPopuli are both from European Parliamentary proceeding scenario. 

\textbf{Found Speech: MiniPS2SLURP.}
We use \textit{SLURP}~\cite{bastianelli2020slurp} as the target domain text-to-semantics data $D^{T \shortrightarrow L, t}$.  \textit{Mini-PS}~\cite{galvez2021people} provides the external-domain speech-to-text pairs $D^{A \shortrightarrow T, \epsilon}$. 
SLURP is in the voice command domain for controlling family robots. 
But Mini-PS is a subset of People's Speech corpus, with 32,255 speech-to-text pairs in diverse domains, such as TV, news, and sermons.
We then mix $D^{A \shortrightarrow T, \epsilon}$ from Mini-PS and $D^{A \shortrightarrow T, t}$ from SLURP for $D^{A \shortrightarrow T}$. 
MiniPS2SLURP is \textit{mismatched} with SLURP's target domain, as it is dominated by \textbf{found} data from other domains.

For fair comparison, in the above two datasets, we provide $D^{A \shortrightarrow L, t} $ that has the same size and speech as  $D^{A \shortrightarrow T, t} $. The $D^{A \shortrightarrow L, t} $ is only used to learn ${\Theta}^{A \shortrightarrow L, t}$ and not applied to learn our $\tilde{\Theta}^{A \shortrightarrow L}$.

We use the full SLURP test set as the test set in MiniPS2SLURP, and half of the dev set in SLUE-VoxPopuli as the test set in VoxPopuli2SLUE.
The dataset statistics, data annotations, and data usages are in Table~\ref{tab:dataset_stats} with sample data in Table~\ref{table:data_example} and domain similarity analysis in Sec.~\ref{sec:domain_similarity}.

\section{Cross-Modal Selective Self-Training}
\label{sec:framework}
\subsection{Introduction of a Basic SLU Model}
\label{sec:slu_model}
Given a sequence of acoustic features $\mathbf{A}$,
the SLU models $\Theta^{A \shortrightarrow L, t} $ and $\tilde{\Theta}^{A \shortrightarrow L}$ extract sentence-level semantics (i.e., intents) and token-level semantics (i.e., entity tags). To support these multiple types of semantic tags, we use a sequence-to-sequence architecture~ \cite{bastianelli2020slurp,ravanelli2021speechbrain}, in which the output is a sequence $\mathbf{Y}$ 
that consists of semantic types with their tags. The SLU model uses a speech encoder to encode $\mathbf{A}$ into a sequence of speech representations, and uses an attentional sequence decoder to generate the output sequence $\mathbf{Y}$. 
The $\Theta^{A \shortrightarrow L, t} $ is trained by loss $\mathcal{L}^{A \shortrightarrow L}$ that maximizes the likelihood of generating the correct semantic sequence given the observation.

\subsection{Overview of our Model: CMSST}
The speech-to-text data $D^{A \shortrightarrow T}$ could provide more resource for SLU training. However, the possible domain mismatch across $D^{T \shortrightarrow L, t}$ and $D^{A \shortrightarrow T, \epsilon}$ can lead to sample noise and label noise. Besides, the imbalance of collected $D^{A \shortrightarrow T}$ may lead to inefficient  model training. 
Thus, we propose a Cross-Modal Selective Self-Training (CMSST) framework to  alleviate the noise and imbalance issue in using  $D^{A \shortrightarrow T}$ and $D^{T \shortrightarrow L, t}$ to learn our E2E SLU model $\tilde{\Theta}^{A \shortrightarrow L}$. We later show in Table~\ref{table:main_results} that CMSST achieves higher performance and efficiency with fewer training samples. 

Figure~\ref{fig:overview}(b) illustrates CMSST. 
First, it computes text similarity to exclude sample pairs in $D^{A \shortrightarrow T}$ that significantly diverge from the text in $D^{T \shortrightarrow L, t}$. 
Second, it takes the distribution of the dataset into consideration, and further filters $D^{A \shortrightarrow T}$ using our novel MCSS to reduce the imbalance within  $D^{A \shortrightarrow T}$ itself. These two steps are described in Sec.~\ref{sec:sample_selection}. 
Lastly, it uses our novel cross-modal selective training method, described in Sec.~\ref{sec:cmsn}, to reduce the impact of noisy labels predicted by an NLU model $\Theta^{T \shortrightarrow L, t}$. The NLU model $\Theta^{T \shortrightarrow L, t}$ is pretrained on $D^{T \shortrightarrow L, t}$.

\subsection{Reducing Sample Noise and Imbalance}
\label{sec:sample_selection}

\textbf{Text similarity based selection.}
The sample selection is firstly performed in a text embedding space. K-Means ~\cite{xu2005survey} is further employed to cluster in the text embedding space for texts from $D^{T \shortrightarrow L, t}$. For each text in $D^{A \shortrightarrow T}$, a text similarity score is defined as the distance to the closest clustering centroid of $D^{T \shortrightarrow L, t}$. Then a threshold based on the text similarity scores is set to exclude $D^{A \shortrightarrow T}$ pairs whose text is dissimilar.

\begin{figure}[t]
\centering
\includegraphics[width=0.48\textwidth]{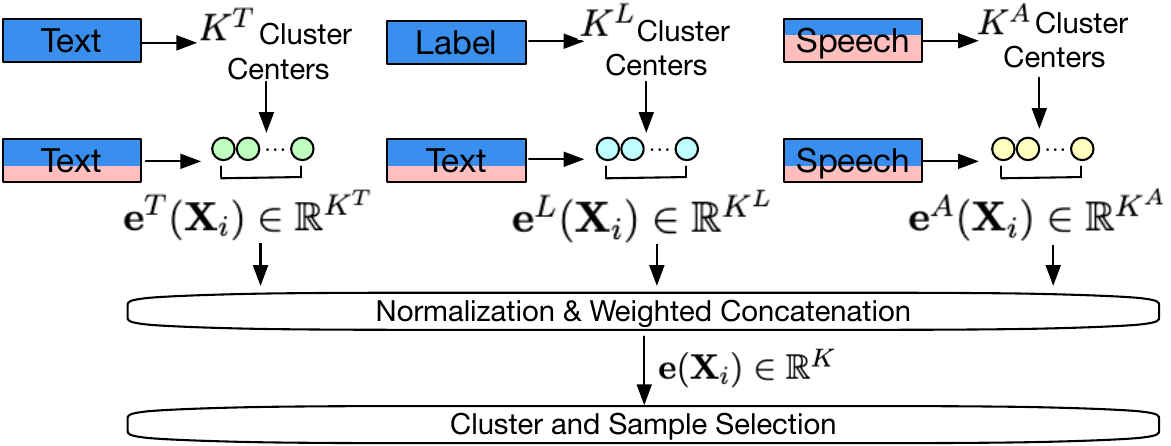}
\caption{
MCSS diagram (detailed in Sec.~\ref{sec:sample_selection}). We use superscripts ${T}$, ${A}$, and ${L}$ to each denote text, speech, and semantic modalities. 
Blue denotes target domain $t$ while pink denotes external domain $\epsilon$. Hence, the blue boxes depict $D^{T \rightarrow L, t}$ data, while blue-pink boxes represent $D^{A \rightarrow T}$ data.
}
\label{fig:MCSS}
\end{figure}

\noindent \textbf{Multi-view Clustering-based Sample Selection (MCSS). } 
Though the above selection process removes speech-text pairs in the mismatched domain, the remaining pairs can still be imbalanced. The imbalanced data distribution introduces bias (e.g., some latent semantic classes may be overrepresented in the found data) into the training and decreases training efficiency. Therefore, it is important to balance the remaining speech-text pairs. 
Since each speech-text pair contains audio, text, and latent semantic information, we propose MCSS to balance these three components. Figure \ref{fig:MCSS} illustrates MCSS's workflow. We use superscripts ${T}$, ${A}$, and ${L}$ to each denote the text, speech, and semantic modalities, respectively.

First, for the text and speech modalities, we use K-Means to cluster texts in $D^{T \shortrightarrow L, t}$ and audio utterances in $D^{A \shortrightarrow T}$. The text embedding is SentenceBERT~\cite{reimers2019sentence} or the average of GloVe~\cite{pennington2014glove}. 
The speech embedding is the average of  HuBERT~\cite{hsu2021hubert}. 
This step outputs $K^{T}$ centroids in the text modality of $D^{T \shortrightarrow L, t}$, and outputs  $K^{A}$ centroids in the audio modality of $D^{A \shortrightarrow T}$.

To represent the semantic space, each entity type in $D^{T \shortrightarrow L, t}$ is an averaged text embedding on all text spans inside that entity type, which is detailed in Sec.~\ref{sec:model_app}. Therefore, the number of entity centroids $K^{L}$ is the number of entity types. We denote these centroids as $\{\mu^{v}_k\}$ for $k \in K^{v}$ and $v \in \{T, A, L\}$ across three modalities.

Given a sample $\mathbf{X}_i$ in $D^{A \shortrightarrow T}$, its distance to the $k$-th clustering centroid $\mu^{v}_k$ in modality $v$ is denoted as $d^{v}(\mathbf{X}_i, \mu^{v}_k)$. Then, we compute the sample's \textit{modality-specific view} $e^{v}(\mathbf{X}_i) \in \mathbb{R}^{K^v}$ as the sample's distances to all centroids in modality $v$,
\begin{equation}
\label{eq:mcss_merge}
  e^{v}(\mathbf{X}_i) = [\cdots, d^{v}(\mathbf{X}_i, \mu^{v}_k), \cdots]
\end{equation}
\noindent and $k\in\{1, 2, ..., K^{v}\}$.

Among three views, $\mathbf{e}^{T}(\mathbf{X}_i)$ and $\mathbf{e}^{L}(\mathbf{X}_i)$ contain information related to $T \shortrightarrow L$ domain, while $\mathbf{e}^{A}(\mathbf{X}_i)$ is generated from speech representation that highly correlates acoustic features in $D^{A\shortrightarrow T}$.

We use cosine distance for all three views (speech, text, and latent semantics, detailed in Sec.~\ref{sec:model_app}). As they are in different scales, we apply Zero-score normalization in each view. In addition, to address the different importance across different views, we use adjustable scalar weight for each view. The multi-view representation is then created by weighted concatenations as $\mathbf{e}(\mathbf{X}_i)=[w^{T}\mathbf{e}^{T}(\mathbf{X}_i), w^{A}\mathbf{e}^{A}(\mathbf{X}_i), w^{L}\mathbf{e}^{L}(\mathbf{X}_i)]$ and $\mathbf{e}(\mathbf{X}_i)\in\mathbb{R}^K$ with $K = K^{T}+K^{A}+K^{L}$. 
The $\mathbf{e}(\mathbf{X}_i)$ is in a joint space of speech, text, and latent semantics, constructed by the $K$ cluster centroids.

To obtain samples that are balanced in this joint space, we then apply the K-Means algorithm on these multi-view representations $\{\mathbf{e}(\mathbf{X}_i)\}$ to get $R$ clusters. Next, we select the equal number of samples for each cluster, and these samples are nearest to the cluster centroid they belong to. Suppose we target for $N$ samples out of the algorithm, then each cluster selects ($\lfloor\frac{N}{R}\rfloor$) of the nearest samples. More details are in Sec.~\ref{sec:model_app}.

\subsection{Reducing Label Noise}
\label{sec:cmsn}
Given the selected speech-to-text pair set $\tilde{D}^{A \shortrightarrow T}$ from MCSS, the pretrained NLU model $\Theta^{T \shortrightarrow L, t}$ predicts pseudolabels. An SLU model is then trained on the speech and its pseudolabels. 
However, these pseudolabels are noisy due to prediction errors in the imperfect NLU model $\Theta^{T \shortrightarrow L, t}$. 
To mitigate label noise, we propose the \textbf{Cross-Modal SelectiveNet (CMSN)} for selective learning. To our best knowledge, we are the first to propose a selective learning method in a cross-modal setting.

\begin{figure}[t]
\centering
\includegraphics[width=0.48\textwidth]{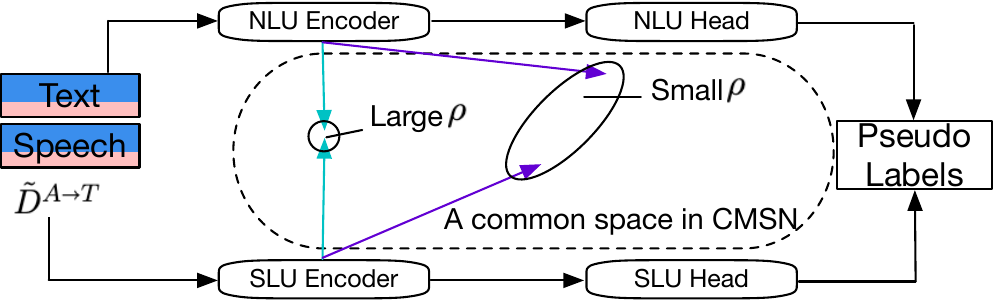}
\caption{Diagram of workflow for CMSN (described in Sec.~\ref{sec:cmsn}), where green or purple arrows are a pair of text and speech. $\rho$ is a selective score described in Eq.~(\ref{eq:selective_score}), where larger $\rho$ indicates projected representations that are more similar.
}
\label{fig:cmsn}
\end{figure}

Figure~\ref{fig:cmsn} illustrates our CMSN. 
For a speech-to-text pair $\mathbf{X}_i$ from $\tilde{D}^{A \shortrightarrow T}$, a text encoder in $\Theta^{T \shortrightarrow L, t}$ and a speech encoder in $\tilde{\Theta}^{A \shortrightarrow L}$ extract their modality-specific embedding vectors $\mathbf{f}^{T}_i$ and $\mathbf{f}^{A}_i$.
Because these embeddings are from the same speech-to-text pair in $\tilde{D}^{A \shortrightarrow T}$, they share a common semantic space. Therefore, we learn modality-specific projections to map the $i$-th sample embeddings to vectors of equal dimension,
\begin{equation}
\mathbf{p}^{v}_{i}=\mathbf{P}^{v} \mathbf{f}^{v}_i, 
\;
\mathbf{q}^{v}_{i}=\mathbf{Q}^{v}\mathbf{f}^{v}_i
\;
\label{eq:text_common1}
\end{equation}
where $v \in \{T, A\}$, $\mathbf{p}$ is in the shared common semantic space, and $\mathbf{q}$ is in the second  common space introduced later. 
We can measure cross-modal loss $\mathcal{L}_{cm1_i}$ by the distance between their common semantic space representations,
\begin{equation}
\mathcal{L}_{cm1_i}=||\mathbf{p}^{T}_{i}-\mathbf{p}^{A}_{i}||.
\label{eq:loss_cm_i}
\end{equation}

To facilitate selective learning, we compute a scalar selective score $\rho\in (0, 1)$ through a selection function $g(\cdot)$ as below,
\begin{equation}
\rho_i=g(\mathbf{p}^{T}_i, \mathbf{p}^{A}_i),
\label{eq:selective_score}
\end{equation}
where $g$ is a multilayer perceptron with a sigmoid function on top of the last layer.
With the selective score, we define the following selective learning loss $\mathcal{L}_{sel}$ that ignores samples with low selection scores,
\begin{eqnarray}
\label{eq:L_sel}
\mathit{\mathcal{L}_{sel}} & = & \alpha \cdot  [\max\left(\tau-E[\rho_i], 0)\right]^2 \\
& + & \beta \cdot \frac{E[\rho_i \mathcal{L}_{cm1_i} + \rho_i \mathcal{L}^{A\shortrightarrow L}]}{E [\rho_i]} \nonumber
\end{eqnarray}
\noindent where $\alpha$ and $\beta$ are scalar weights. 
The first term in Eq. (\ref{eq:L_sel}) has a hyper-parameter $\tau\in[0, 1]$, which is defined as the target coverage in \citet{geifman2019selectivenet}. 
Concretely, the first term encourages the selective network to output selective scores closer to $\tau$, especially if the selective scores are small at the beginning of model training. 

For the Eq.~(\ref{eq:L_sel}) second term, we weigh both $\mathcal{L}_{cm1_i}$ and $\mathcal{L}^{A \shortrightarrow L}$ by $\rho_i$. This is because certain text embeddings could be inaccurate, which can make the $\mathcal{L}_{cm1_i}$ large, and the pseudolabel derived from the text embedding becomes noisy, indicating its $\mathcal{L}^{A \shortrightarrow L}$ needs to be down-weighted. In this case, if $\mathcal{L}_{cm1_i}$ is large, the Eq.~(\ref{eq:L_sel}) second term encourages a smaller $\rho_i$ from Eq. (\ref{eq:selective_score}). A reduced $\rho_i$ mitigates the impact of $\mathcal{L}^{A \shortrightarrow L}$, thus enpowering CMSN to selectively trust $\mathcal{L}^{A \shortrightarrow L}$. 
The final loss is,
\begin{equation}
\mathcal{L}= \mathcal{L}^{A \shortrightarrow L} +  \mathcal{L}_{sel} + \gamma \mathcal{L}_{cm_2}
\label{eq:L_final}
\end{equation}
where $\gamma$ is the weight of auxiliary cross-modal loss $\mathcal{L}_{cm_2}$. The $\mathcal{L}_{cm_2}$ encourages the common space learning by the expectation (mean) of all sample cross-modal differences weighted by respective $\rho$, 
\begin{eqnarray}
\label{eq:L_aux_cm2}
\mathcal{L}_{cm2} = E[\rho_i||\mathbf{q}^{T}_i - \mathbf{q}^{A}_i||]
\end{eqnarray}

The use of the $\mathcal{L}_{cm2}$ via another projection $\mathbf{Q}^{v}$ 
is essential to optimize the
selective network~\cite{geifman2019selectivenet}. With $\mathcal{L}_{cm2}$, the selective network can additionally learn the alignment of cross-modal features. 
Therefore, $\mathcal{L}_{cm2}$ avoids overfitting the selective network to the biased subset, before accurately learning low-level speech features. 

\section{Experiments}
\label{sec:experiments}
We now compare our proposed framework to baselines on the two datasets introduced in Sec.~\ref{sec:prbolem_set}.

\subsection{Performance Metrics} 
Following~\citet{bastianelli2020slurp}, we report (1) sentence-level classification performance using average accuracy (\textbf{Acc.}) on classifying Scenario ({Scenario Acc.}), action ({Action Acc.}) and intent ({Intent Acc.}), and (2) NER performance from the list of entity type-value pairs. The \textbf{Entity-F1} is a sentence-level NER metric, in which the correctness of entity type-value pairs and their appearance orders are measured. \textbf{Word-F1} drops the penalty on their appearance orders. \textbf{Char-F1} further relaxes exact match at word level and allows character-level match of entity values. 
To measure the training efficiency, we report numbers of used speech-text pairs (sum of $\|D^{A \shortrightarrow T, t}\|$ and $\|D^{A \shortrightarrow T, \epsilon}\|$) and training time. Experiments were run on a single GPU 3090 with 24G memory.

\begin{table*}[tbh]
\small
\begin{tabular}{l|r|rrr|r|rrr|r}
\hline
\multirow{2}{*}{\textbf{Models}}& \multirow{2}{*}{\scriptsize{$\|D^{A \shortrightarrow L, t} \|$}}& \multirow{2}{*}{\textbf{\scriptsize{$\|D^{A \shortrightarrow T, t}\|$}}}  & \multirow{2}{*}{\textbf{\scriptsize{$\|D^{A \shortrightarrow T, \epsilon}\|$}}} & \multirow{2}{*}{$N$ $\downarrow$} & \textbf{Acc. $\uparrow$} & \multicolumn{3}{c|}{\textbf{NER F1} (in \%) $\uparrow$} & \textbf{Time $\downarrow$} \\
 &  &  &  &   & (in \%) & Entity     & Word         & Char & (in hrs) \\
\hline
\hline
\multicolumn{9}{l}{\textit{MiniPS2SLURP} (Found Speech)} \\
\hline
Target model $\Theta^{A \shortrightarrow L, t} $ & 22.8k & N/A & N/A & N/A & 76.0 &40.9 & 51.7 & 55.8 & 16 \\ 
\hline
$\tilde{\Theta}_{Full}^{A \shortrightarrow L}$~\cite{pasad-etal-2022-use} & 0 & 22.8k & 32.3k &55.1k & 74.9 & 34.9 & 48.8 & 52.0  & 43
\\ 
$\tilde{\Theta}_{RSamp}^{A \shortrightarrow L}$ & 0 & 14.4k & 20.6k & \bf{35k} & 73.5 & 33.9	& 47.5 &50.9 & \bf{27} \\ 
Our $\tilde{\Theta}^{A \shortrightarrow L}$ (GloVe)   & 0 & 21.6k  & 13.4k & \bf{35k} & {75.2} & {34.9} & {48.8} & {52.2} & \bf{27}  \\
Our $\tilde{\Theta}^{A \shortrightarrow L}$ (SentBERT)  & 0 & 22.1k  & 12.9k & \bf{35k} & \textbf{75.4} & \textbf{35.7} & \textbf{49.3} & \textbf{52.9} & \bf{27}  \\
\hline
\hline   
\multicolumn{9}{l}{\textit{VoxPopuli2SLUE} (Matched Speech)} \\
\hline
Target model $\Theta^{A \shortrightarrow L, t} $  & 2,250 & N/A & N/A & N/A & N/A & 36.0          & 45.2           & 47.7 & 2 \\
\hline
$\tilde{\Theta}_{Full}^{A \shortrightarrow L}$~\cite{pasad-etal-2022-use} & 0 & 2,250 & 182.5k &184.8k & N/A & 37.0 & \textbf{50.3} & \textbf{53.9}  & 225 \\
$\tilde{\Theta}_{RSamp}^{A \shortrightarrow L}$ & 0 & 68 & 5.6k & 5.6k &N/A & 35.7 &	47.8 &	50.5  & \bf{6} \\
Our $\tilde{\Theta}^{A \shortrightarrow L}$ (GloVe)   & 0 & 59 & 5.5k & \bf{5.5k} & 
N/A& {36.8} & {49.0} & {52.3}  & \bf{6} \\  
Our $\tilde{\Theta}^{A \shortrightarrow L}$ (SentBERT)   & 0 & 61 & 5.5k & \bf{5.5k} &
N/A& \textbf{38.0} & {49.3} & {52.4}  & \bf{6} \\  
\hline
\end{tabular}
\caption{Comparison between our  proposed CMSST and baselines. The selected speech-text pairs size $N$ is the sum of $\|D^{A \shortrightarrow T, t}\|$ and $\|D^{A \shortrightarrow T, \epsilon}\|$.
Our model utilizes \textbf{significantly fewer speech-text pairs} and \textbf{training time} compared with $\tilde{\Theta}_{Full}^{A \shortrightarrow L}$ (which uses all speech-text pairs), yet achieves \textbf{comparable or superior accuracy and F1 scores}. 
}
\label{table:main_results}
\end{table*}

\subsection{Baselines \& Experiment Setups}
\label{sec:baselines_experiment_setups}
We compare our method with two types of methods: (1) a strong baseline that uses all of the ASR data~\cite{pasad-etal-2022-use}, denoted as $\tilde{\Theta}_{Full}^{A \shortrightarrow L}$ and  
(2) a model that random samples training data to have data size comparable to our method, denoted as $\tilde{\Theta}_{RSamp}^{A \shortrightarrow L}$~\footnote{We forego comparisons with ~\citet{mdhaffar2022e2e}, due to its unreleased code and use of "pseudospeech"-semantics pairs, in contrast to our use of speech-"pseudosemantics" pairs like~\citet{pasad-etal-2022-use}.}. We also report the performance of $\Theta^{A \shortrightarrow L, t} $ that is trained with target domain speech-to-semantics data $D^{A \shortrightarrow L, t}$. 
We compare text-similarity selection by GloVe and SentenceBERT (Abbr: SentBERT). 

Our data split details and sample data are provided in Sec.~\ref{app:data_example}. Additionally, more implementation details of our experiments can be found in Sec.~\ref{app:implementation_details}.

\subsection{Main Results}
\label{sec:main_results}

The main results of the proposed model on the two datasets are illustrated in Table~\ref{table:main_results}. 
Firstly, our proposed method using SentBERT embedding can surpass the strong baseline $\tilde{\Theta}_{Full}^{A \shortrightarrow L}$ that uses all training samples
in both GloVe-based and SentBERT-based text-similarity. 
For example, on the NER task, our SentBERT-based model achieved an entity-F1 score of 38.0\% on the matched speech VoxPopuli2SLUE dataset, surpassing the full system, which scored 37.0\%. Besides, our method shows a significant reduction of training time from 225 hours to 6 hours and number of speech-text pairs from 182k to 5k, 
as our method uses 2.7\% of the full dataset size. 
On the found speech MiniPS2SLURP, our SentBERT-based model achieves higher performance in both accuracy and F1 scores and higher training efficiency. For example, it improves 1.2 points in Entity F1 than $\tilde{\Theta}_{Full}^{A \shortrightarrow L}$ that uses 1.5 times of training time and data size of ours.

Our performance gain is apparent when compared to $\tilde{\Theta}_{RSamp}^{A \shortrightarrow L}$, using a similar size of randomly sampled training data. In such a case, entity F1 scores on two datasets drop by around 1 and 2 percents compared to our GloVe-based and SentBERT-based methods, respectively. 

The proposed method surpasses the performance of the target model $\Theta^{A \shortrightarrow L, t} $ in the matched speech VoxPopuli2SLUE set. For instance, our SentBERT-based model has word-level entity F1 improved to 49.3\% from 45.2\% of the target model. On the found speech MiniPS2SLURP, the difference to the target model is reduced to 0.6\% by our method, compared to 1.1\% by $\tilde{\Theta}_{Full}^{A \shortrightarrow L}$ and 2.5\% by $\tilde{\Theta}_{RSamp}^{A \shortrightarrow L}$ in terms of Acc. 

The results on SentBERT-based text-similarity marginally perform better than the GloVe-based. Except the 1.2 percents difference on NER F1 on VoxPopuli2SLUE, all the other metrics on both two datasets show less than 1 percent difference. The marginal difference between two methods is similar to other self-training work~\cite{du2020self}. Due to the slight difference, our ablation studies use GloVe-based text selection for faster speed.

\begin{table*}[tbh]
\small
\begin{tabular}{l|r|rrr|r|rrr|r}
\hline
\multirow{2}{*}{\textbf{Models}}& \multirow{2}{*}{\scriptsize{$\|D^{A \shortrightarrow L, t} \|$}}& \multirow{2}{*}{\textbf{\scriptsize{$\|D^{A \shortrightarrow T, t}\|$}}}  & \multirow{2}{*}{\textbf{\scriptsize{$\|D^{A \shortrightarrow T, \epsilon}\|$}}} & \multirow{2}{*}{$N$ $\downarrow$} & \textbf{Acc. $\uparrow$} & \multicolumn{3}{c|}{\textbf{NER F1} (in \%) $\uparrow$} & \textbf{Time $\downarrow$} \\
 &  &  &  &   & (in \%) & Entity     & Word         & Char & (in hrs) \\
\hline
\hline
\multicolumn{9}{l}{\textit{MiniPS2SLURP} (Found Speech)} \\
\hline
Target model $\Theta^{A \shortrightarrow L, t} $ & 22.8k & N/A & N/A & N/A & 76.0 &40.9 & 51.7 & 55.8 & 16 \\ 
\hline
Only SLURP data   & 0 & 22.8k  & 0 & \bf{22.8k} & {70.0} & {30.8} & {44.1} & {46.9} & \bf{14}  \\
Only Mini-PS data  & 0 & 0 & 32.3k & {32.3k} & 16.8 & 10.3	& 18.1 &18.8 & {25} \\ 
Full data ($\tilde{\Theta}_{Full}^{A \shortrightarrow L}$) & 0 & 22.8k & 32.3k &55.1k & 74.9 & 34.9 & 48.8 & 52.0  & 43
\\ 
Our $\tilde{\Theta}^{A \shortrightarrow L}$ (SentBERT)  & 0 & 22.1k  & 12.9k & {35k} & \textbf{75.4} & \textbf{35.7} & \textbf{49.3} & \textbf{52.9} & {27}  \\
\hline
\end{tabular}
\caption{Comparison among different methods in using only SLURP data, only Mini-PS, full data, and our model in MiniPS2SLURP. The selected speech-text pairs size $N$ is the sum of $\|D^{A \shortrightarrow T, t}\|$ and $\|D^{A \shortrightarrow T, \epsilon}\|$.}
\label{table:only_slurp_only_people}
\end{table*}

\begin{table*}[]
\small
\centering
\begin{tabular}{l|ccc|c}
\hline
   & Scenario Acc. & Action Acc. & Intent Acc. & Mean Acc. \\
   \hline
Target model      & 80.7          & 74.7        & 72.6        & 76.0      \\
\hline
$\tilde{\Theta}_{Full}^{A \shortrightarrow L}$~\cite{pasad-etal-2022-use} & 80.1          & 73.4        & 71.2        & 74.9      \\
$\tilde{\Theta}_{RSamp}^{A \shortrightarrow L}$  & 79.0          & 71.9        & 69.6        & 73.5      \\
Our model $\tilde{\Theta}^{A \shortrightarrow L}$(GloVe) & 80.6          & 73.5        & 71.4        & 75.2      \\
Our model $\tilde{\Theta}^{A \shortrightarrow L}$ (SentBERT)                                         & \bf80.7          & \bf73.9        & \bf71.6        & \bf75.4     \\
\hline
\end{tabular}
\caption{More details about each dimension of accuracy in MiniPS2SLURP. The mean accuracy is aligned with Table~\ref{table:main_results}.}
\label{table:slurp_detail_acc}
\end{table*}

\section{Analysis}
\label{sec:analysis}
\subsection{Ablation Studies}
\label{sec:ablations}
\begin{figure}[!bth]
\centering
\includegraphics[width=0.42\textwidth]{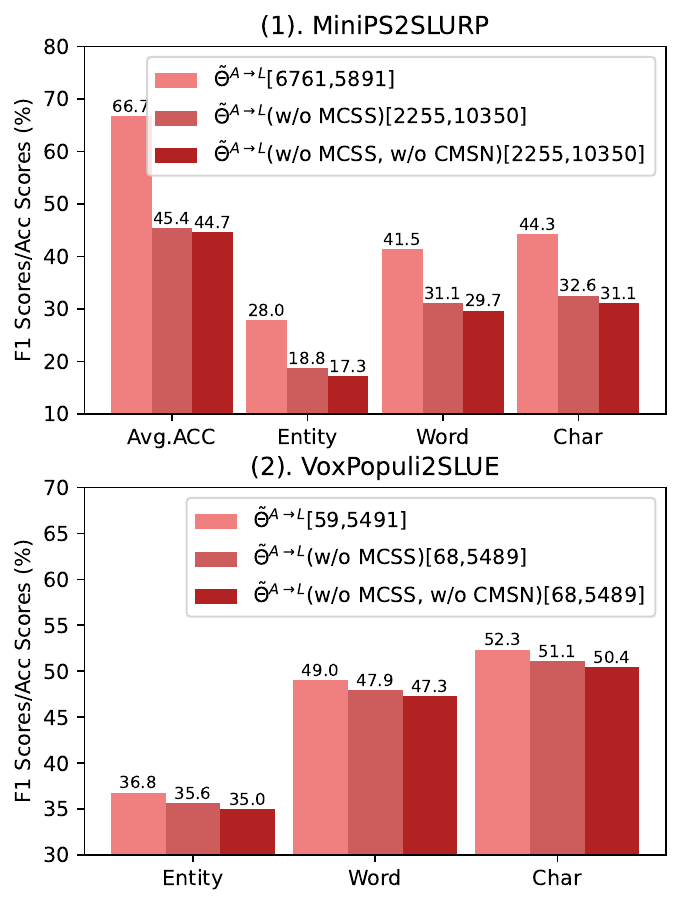}
\caption{
Ablation study on the effectiveness of multi-view sample selection and selective training on $\tilde{\Theta}^{A \shortrightarrow L}$. The pseudolabels are from BERT-based $\Theta^{T \shortrightarrow L, t}$. Their {$\|D^{A \shortrightarrow T, t}\|$} and {$\|D^{A \shortrightarrow T, \epsilon}\|$} size are each listed in square brackets for each configuration. The selection size $N$ is 12.6k and 5.5k for the two datasets respectively.
}
\label{fig:ablation_study_bert}
\end{figure}

\textbf{Multi-view Clustering-based Sample Selection (MCSS).} 
We use different thresholds on the text similarity scores and control the selective size $N$ to be approximately the same for a fair comparison. Results are shown in Figure~\ref{fig:ablation_study_bert}.
On the found speech MiniPS2SLURP, we use its subset for the ablation study and observe that removing MCSS (\textbf{w/o MCSS}) hurts performance. 
For example, using MCSS, entity F1 score is improved from 18.8\% to 28.0\%, a 49\% relative improvement.  
Another observation is that MCSS apparently has fewer external-domain samples than without using the MCSS algorithm. For instance, w/o MCSS, the {$\|D^{A \shortrightarrow T, \epsilon}\|=10350$}, which is almost twice as large as {$\|D^{A \shortrightarrow T, \epsilon}\|=5891$} with MCSS in $\tilde{\Theta}^{A \shortrightarrow L}$.

\noindent \textbf{Cross Modal SelectiveNet (CMSN).}
Results in Figure~\ref{fig:ablation_study_bert} show that further removing selective training (w/o MCSS, w/o CMSN) results in performance loss. On MiniPS2SLURP, the entity F1 score is improved from 17.3\% to 18.8\% if using CMSN, a relative 8.7\% improvement. 

Performance improvements are also observed for the matched speech VoxPopuli2SLUE dataset in Figure~\ref{fig:ablation_study_bert}. 
These results show that both reducing imbalance by sample selection (MCSS) and reducing label noise by selective learning (CMSN) can improve performance by the proposed framework.

\subsection{Impacts from NLU Backbone} 

\begin{table}[!htb]
\small
\begin{tabular}{l|c|lll}
\hline
\multirow{2}{*}{\textbf{Backbone}} & \multirow{2}{*}{\textbf{MCSS+CMSN}}  &
\multicolumn{3}{c}{\textbf{NER F1} (in \%)} \\      
& & Entity  & Word & Char \\
\hline
\multirow{2}{*}{LSTM} &
 & 
35.1                & 45.5 & 48.6          \\
& \checkmark 
& \bf{36.6} &	\bf{46.4}	 &\bf{49.1} \\
\hline
\multirow{2}{*}{BERT} &         
 & 35.0     & 47.3 & 50.4           \\
& \checkmark 
& \bf{36.8} & \bf{49.0}  & \bf{52.3} \\ 
\hline   
\end{tabular}
\caption{Impact comparison of using LSTM and BERT NLU backbones, on VoxPopuli2SLUE. Both backbones have $\|D^{A \shortrightarrow T, t}\|=68$ and $\|D^{A \shortrightarrow T, \epsilon}\|=5489$ after text similarity based selection and MCSS.}
\label{table:main_slue_lstm1}
\end{table}

In this section, we conduct experiments on VoxPopuli2SLUE to study the impact of different NLU backbones in $\Theta^{T\shortrightarrow L, t}$. The comparison reveals the effectiveness of the proposed framework in dealing with different qualities of pseudolabels. 
We select LSTM and BERT due to their wide applications. The BERT-based backbone was fine-tuned from pretrained ``bert-base-uncased''. We fix its encoder but train prediction heads. The LSTM backbone was trained from scratch. Both backbones are trained from 2250 samples in $D^{T \shortrightarrow L, t}$. We measure their performance on the test set using ground truths from their text inputs. The BERT-based NLU backbone has higher NER performance than the LSTM-based NLU backbone, with 39.3\% vs. 36.7\% entity F1 Score (not listed in tables).

From Table~\ref{table:main_slue_lstm1}, 
we observe that (1) labels from a BERT-based backbone result in comparable or higher performance, (2) using the framework (i.e., w/ MCSS+CMSN checked) consistently improves the performances of the learned SLU models.

\subsection{Sample Diversity}
\label{sec:diversity}

\begin{table}[tb]
\scriptsize
\begin{tabular}{l|ll|lll}
\hline
\textbf{Sampling}   & \multirow{2}{*}{\textbf{\scriptsize{$\|D^{A \shortrightarrow T, t}\|$}}} &\multirow{2}{*}{\textbf{{\scriptsize{$\|D^{A \shortrightarrow T, \epsilon}\|$}}}}    & \multicolumn{3}{c}{\textbf{\scriptsize{Diversity (Entropy)}}} \\      
\textbf{Method}& & & $T$  & $L$ & $A$  \\
\hline
Equal & 59 & 5,491  & \bf{3.94} & \bf{1.34}   & \bf{4.36} \\  
Random & 61 &	5,495 & 3.84 & 1.24  & 4.34 \\  
Extreme  & 47 &	5,509 & 3.78 & 1.20  & 2.55 \\
\hline
w/o MCSS & 68 & 5,489 &2.75 & 1.03  &4.27
\\
\hline
\end{tabular}
\caption{Sample diversity from views of the three modalities (text (T), semantic labels (L), and audio (A)). They are computed as entropy on samples from different selection methods. Results are on VoxPopuli2SLUE.}
\label{table:entropy_three_view}
\end{table}

This section provides further analysis of MCSS. The observation in Figure~\ref{fig:ablation_study_bert} shows improved performance and increased proportions of in-domain data. Our hypothesis is that samples are more diverse due to the sample selection method described in Sec.~\ref{sec:sample_selection}. To quantify this, we measure the entropy of the selected samples, specifically for each view $v \in \{T, L, A\}$. Entropy in each view $v$ is computed as $ -\sum_{k=1}^{K^{v}} \frac{n^{v}_k}{N}\log \frac{n^{v}_k}{N},
$
where $K^{v}$ is the number of clusters for view $v$, $n^{v}_k$ is the number of samples in cluster $k$ for view $v$, and $N$ is the total number of samples. 
Their results are in Table~\ref{table:entropy_three_view}. For comparison, we also measure the entropy from random sampling (Random) and entropy from selecting samples with as few clusters as possible (Extreme). 
We observe that the entropy from the equal sampling method is larger than random sampling in all three views. The extreme sampling method has the lowest entropy, compared to the other two sampling methods. As a larger entropy indicates more diversity, we conclude that our equal sampling results in the largest diversity among these methods. We also list the entropy on a similar size of filtered samples without MCSS; their entropies in three views are much lower compared to our equal sampling method.

\subsection{More Analysis of SLURP Results}
\label{sec:more_ana_res_slurp}
Table~\ref{table:only_slurp_only_people} presents the results obtained by solely using SLURP data and only using Mini-PS data. From the table, it is evident that exclusively relying on Mini-PS data results in significantly poorer performance, due to a domain mismatch between SLURP and Mini-PS datasets. Furthermore, the analysis suggests that only a portion of the Mini-PS data contributes positively to learning on the MiniPS2SLURP dataset. 

Additionally, we provide accuracy breakdowns across various dimensions (e.g., Scenario, Action, and Intent) in Table~\ref{table:slurp_detail_acc}. Analysis of the table reveals that our model (SentBERT) consistently outperforms other baselines across each dimension.

\subsection{Parameter Analysis \& Other Experiments}
\label{sec:analysis_tau}
\begin{figure}[!tbp]
\centering
\includegraphics[width=0.44\textwidth]{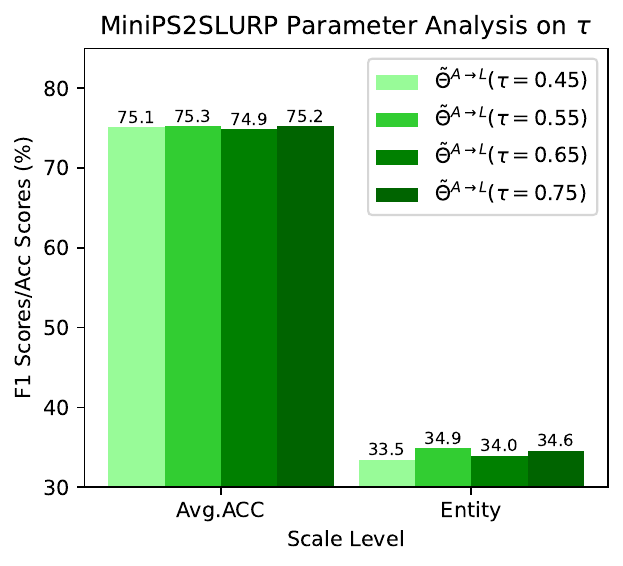}
\caption{Entity F1 Scores and Acc. on the found speech MiniPS2SLURP dataset, where all groups have the same $\|D^{A \shortrightarrow T, t}\|=21597$ and $\|D^{A \shortrightarrow T, \epsilon}\|=13400$. 
} 
\label{fig:analysis_tau}
\end{figure}

Figure~\ref{fig:analysis_tau} shows Entity F1 scores and average accuracy on MiniPS2SLURP. The pseudolabels are from the BERT-based $\Theta^{T \shortrightarrow L, t}$. We observe an optimal value of $\tau=0.55$. Other parameter analysis results in both MCSS and CMSN are in Sec.~\ref{sec:parameters}. Another cluster method and cluster quality are analyzed in Sec.\ref{sec:cluster_quality_analysis}. Examples of output produced by our approach under the ablation setting are presented in Sec.~\ref{app:case_study}.

\section{Conclusion}
To advance zero-shot E2E SLU research, we create two datasets: VoxPopuli2SLUE and MiniPS2SLURP, catering to matched and found speech, respectively. In addition, our framework CMSST tackles the noise and imbalance issues that have been disregarded in previous works. CMSST incorporates MCSS, a method that selects speech-text pairs to simultaneously enhance the diversity of acoustic, text, and semantics, thus addressing the imbalance. Furthermore, CMSN is proposed to mitigate the impact of low-confidence pseudolabels, thereby alleviating the effects of label noise. Extensive experiments on both datasets demonstrated the effectiveness and efficacy of our framework.

\section{Ethical Consideration}
This study pioneers the use of text-semantics and audio-text pairs to learn a SLU model in a zero-shot way. Additionally, we have innovatively addressed issues of noise and imbalance through the implementation of selective self-training methods.

Our research exclusively employs datasets that are publicly available, ensuring transparency and accessibility. The datasets integral to our work are utilized in adherence to their respective licenses, which is verified in Sec. \ref{sec:license}.

All of our used datasets do not have personal identification information. We recommend that any future expansion of this research into areas involving personal or sensitive data should be approached with stringent ethical guidelines in place.

\section{Limitations}
This paper proposes CMSST for zero-shot end-to-end SLU. CMSST has a main limitation in MCSS. Concretely, MCSS has the limitation that the samples are selected from the nearest cluster centers. Alternatively, we can improve MCSS by choosing samples that maximize the mutual information in each cluster, which we leave for future work.

\bibliography{main}

\clearpage
\appendix

\section{Appendix}
\subsection{Domain Similarity Analysis in VoxPopuli2SLUE \& MiniPS2SLURP}
\label{sec:domain_similarity}
\textbf{Analysis of domain similarity.} In discussing domain similarity, it is essential to clarify the ``domain,'' which refers to data collection scenarios in this paper. Each dataset encompasses two domains: the target domain and the external domain. For MiniPS2SLURP, the external domain is an OOD domain, whereas in VoxPopuli2SLUE, the external domain aligns with the target domain. To assess domain similarity, we employ the \textbf{Maximum Mean Discrepancy} (\textbf{MMD})~\cite{wang2020rethink}, a statistical measure gauging differences between two distributions. A MMD value approaching zero indicates closeness between the two distributions. To delve into vocabulary divergence, we measured MMD using the TF-IDF feature, termed \textbf{MMD-TFIDF}. Similarly, to understand semantic divergence, we used the SentenceBERT feature to calculate MMD, which is written as \textbf{MMD-SentBERT}. The results for both datasets are documented in Table~\ref{tab:domain_sim_two_proposed_datasets}. From the table, MiniPS2SLURP exhibits a significant domain divergence between Mini-PS and SLURP, with both MMD-TFIDF and MMD-SentBERT values surpassing 0.6. In contrast, VoxPopuli2SLUE shows minimal divergence, as evidenced by both MMD values being around 0.05—attributable to its external domain being the same as the target domain.

\begin{table}[!bth]
\scriptsize
\centering
\begin{tabular}{l|c|c}
\hline
               & \multicolumn{1}{l|}{MMD-TFIDF $\downarrow$} & \multicolumn{1}{l}{MMD-SentBERT $\downarrow$} \\ \hline
MiniPS2SLURP   & 0.6381                           & 0.6326                              \\ \hline
VoxPopuli2SLUE & 0.0663                           & 0.0416                              \\ \hline
\end{tabular}
\caption{The domain similarity between the target domain and the external domain of the two proposed datasets.}
\label{tab:domain_sim_two_proposed_datasets}
\end{table}

\subsection{Another Cluster Method \& Cluster Quality Analysis}
\label{sec:cluster_quality_analysis}
\textbf{Cluster quality metrics.} For cluster quality metrics, such metrics are typically based on one label per ground-truth sample. However, only MiniPS2SLURP provides these utterance-level labels (e.g., scenarios), while VoxPopuli2SLUE offers only entity-level labels. As a result, we measured the cluster quality only for MiniPS2SLURP. We used two metrics:
\\
(a) \textbf{Purity}~\cite{marutho2018determination}: This metric assigns the majority sample label within a cluster as the cluster's label. The purity is then calculated as the average accuracy across all samples.
\\
(b) \textbf{Normalized Mutual Information (NMI)}~\cite{huang2010shrink}: This metric measures the similarity between two sets of clusters, regardless of potential variations in the number of clusters in each set. In our work, we use NMI to measure the similarity between the ground-truth class labels and cluster results, where each cluster uses the majority sample label within the cluster as its label.

\textbf{Analysis of cluster quality of two cluster methods.} Due to our dataset constraints, where the audio data comes with transcripts but lacks labels in our zero-shot setting, it is inapplicable to measure its clustering. Thus, we can only detail the quality on texts in $D^{T\rightarrow L}$
for two clustering methods, which is shown in Table~\ref{tab:quality_two_datasets}. We additionally experimented with \textbf{hierarchical agglomerative clustering} (abbreviated as \textbf{Hierarchical})~\cite{mullner2011modern}, which recursively merges cluster pairs in the sample data. Table~\ref{tab:quality_two_datasets} reveals a high purity for the clusters, suggesting a dominant presence of samples with consistent labels in each cluster. The high NMI scores further underscore that our clustering aligns closely with the ground-truth labels. Therefore, our chosen clustering techniques, including K-Means and hierarchical agglomerative clustering, exhibit high quality.

\begin{table}[!bht]
\centering
\small
\begin{tabular}{l|r|r}
\hline
         & \multicolumn{1}{l|}{K-Means} & \multicolumn{1}{l}{Hierarchical} \\ \hline
Purity $\uparrow$ & 0.8498                      & 0.8363                            \\ \hline
NMI $\uparrow$    & 0.6307                      & 0.6183                            \\ \hline
\end{tabular}
\caption{The clustering quality of both K-Means and hierarchical agglomerative  clustering on MiniPS2SLURP texts in text-to-semantics pairs.}
\label{tab:quality_two_datasets}
\end{table}

\textbf{Analysis of SLU model performance by two cluster methods}. For the downstream SLU training performance using hierarchical clustering, results are provided in Table~\ref{tab:slu_results_another_cluster}. 
From the table, it is evident that our model, utilizing SentBERT text embedding with hierarchical agglomerative clustering, consistently achieves competitive results, outperforming the random baseline in Table~\ref{table:main_results}. Moreover, in Table~\ref{tab:slu_results_another_cluster}, our model requires significantly fewer samples to achieve an improvement of 1.0 and 1.3 points in average accuracy over the baseline using full samples for MiniPS2SLURP and VoxPopuli2SLUE in Table~\ref{table:main_results}, respectively. This performance improvement shows our model's adaptability to another clustering method.

\begin{table*}[tbh]
\scriptsize
\begin{tabular}{l|r|rrr|r|rrr|r}
\hline
\multirow{2}{*}{\textbf{Models}}& \multirow{2}{*}{\scriptsize{$\|D^{A \shortrightarrow L, t} \|$}}& \multirow{2}{*}{\textbf{\scriptsize{$\|D^{A \shortrightarrow T, t}\|$}}}  & \multirow{2}{*}{\textbf{\scriptsize{$\|D^{A \shortrightarrow T, \epsilon}\|$}}} & \multirow{2}{*}{$N$ $\downarrow$} & \textbf{Acc. $\uparrow$} & \multicolumn{3}{c|}{\textbf{NER F1} (in \%) $\uparrow$} & \textbf{Time $\downarrow$} \\
 &  &  &  &   & (in \%) & Entity     & Word         & Char & (in hrs) \\
\hline
\hline
\multicolumn{9}{l}{\textit{MiniPS2SLURP}(Found Speech)} \\
Our $\tilde{\Theta}^{A \shortrightarrow L}$ (SentBERT, K-Means)  & 0 & 22.1k   & 12.9k & 35k & 75.4 & \textbf{35.7} & \textbf{49.3} & \textbf{52.9} & 27  \\
Our $\tilde{\Theta}^{A \shortrightarrow L}$ (SentBERT, Hierarchical) &0	&22.1k	&12.9k	& 35k &\bf{75.9}	&34.9	&48.9	&52.5	&27 
\\
\hline
\hline   
\multicolumn{9}{l}{\textit{VoxPopuli2SLUE} (Matched Speech)} \\
\hline
Our $\tilde{\Theta}^{A \shortrightarrow L}$ (SentBERT, K-Means)   & 0 & 61 & 5.5k & 5.5k & N/A& 38.0 & {49.3} & {52.4}  & 6 \\  
Our $\tilde{\Theta}^{A \shortrightarrow L}$ (SentBERT, Hierarchical) &0	&61	&5.5k & 5.5k	&N/A	&\bf{38.3}	&48.9	&51.4	&6
\\
\hline
\end{tabular}
\caption{Comparison between K-Means and hierarchical agglomerative clustering (abbreviated as Hierarchical) on the datasets. 
}
\label{tab:slu_results_another_cluster}
\end{table*}

\textbf{Analysis of alignments between the target-domain samples and our selected samples.} In evaluating the alignment results from our data selection, we employed two metrics: 
(1) MMD-TFIDF and (2) MMD-SentBERT. These statistics are detailed in Table~\ref{tab:result_data_selection}. Notably, in the MiniPS2SLURP dataset, our methods produced improved (smaller) values for both MMD metrics compared to full and random baselines. For the VoxPopuli2SLUE dataset, our methods resulted in improved (smaller) values for MMD-TFIDF and similar values for MMD-SentBERT. This suggests that the texts selected using our approach are more aligned, exhibiting less divergence from the target domain in both vocabulary and semantics, underscoring our method's efficacy.

\begin{table}[]
\centering
\small
\begin{tabular}{l|c|c}
\hline
 & MMD-TFIDF $\downarrow$ & MMD-SentBERT $\downarrow$ \\ 
\hline
\hline
\multicolumn{3}{l}{\textit{MiniPS2SLURP}}
\\
\hline
Full                   & 0.0590    & 0.1180       \\ 
Random                 & 0.0589    & 0.1432       \\ 
Ours (Glove)           & 0.0394    & 0.0548       \\ 
Ours (SentBERT)         & \bf{0.0385}    & \bf{0.0539}       \\ 
\hline
\hline
\multicolumn{3}{l}{\textit{VoxPopuli2SLUE}}
\\
\hline
Full                   & 0.0995    & 0.0405       \\ 
Random                 & 0.0653    & \bf{0.0401}       \\ 
Ours (Glove)           & \bf{0.0336}    & 0.0411       \\ 
Ours (SentBERT)         & 0.0403    & 0.0452       \\ \hline
\end{tabular}
\caption{Alignment analysis of data selection results across two datasets. The MMD-TFIDF and MMD-SentBERT are compared to the respective target domain in terms of word frequency and SentBERT embedding.  The method organization mirrors that in Table~\ref{table:main_results} of the manuscript. 
}
\label{tab:result_data_selection}
\end{table}

\subsection{Model}
\label{sec:model_app}
\textbf{Semantic representations.} Specifically, the semantics in $D^{T \shortrightarrow L, t}$ has $K^{L}$ types (i.e. ``LOC'', ``DATE''). 
We build type centroids by using the average GloVe or SentenceBERT features of all slot texts from a semantic type. Consequently, we obtain $K^{L}$ clustering centroids for semantics. For example, suppose we have three entity types: “Date”, “Loc”, and “Person”, provided in $D^{T\rightarrow L, t}$. For the “Date” type, we aggregate all its date labels and then compute the average of the text embeddings of these labels. This average serves as the “Date” entity centroid. Following this process, given the three entity types in this example, we would produce three entity centroids corresponding to “Date”, “Loc”, and “Person”.
\\
\textbf{Detailed explanation of Eq.~\ref{eq:mcss_merge}} In Eq.~\ref{eq:mcss_merge}, $d^T(X_i, \mu^T_k)$ refers to the distance between the embedding of the $i$-th text from speech-text pairs and the $k$-th text cluster center embedding from text clusters in text-semantics pairs.

Also, $d^L(X_i, \mu^L_k)$ refers to the distance between the embedding of the $i$-th text from speech-text pairs and the $k$-th semantic entity centroid embedding from text-semantics pairs (no need for cluster operation).

Finally, $d^A(X_i, \mu^A_k)$ represents the distance between the embedding of the $i$-th speech-text pair's speech and the embedding of the $k$-th speech cluster center from the speech clusters in the speech-text pairs.
The above three processes are also drawn in Fig.~\ref{fig:MCSS}.
\\
\textbf{Normalization methods.}
For the normalization, we use the z-score normalization for ${e}^{v}(\mathbf{X}_i)$, where $v\in\{T, A, L\}$.
After the normalization, the feature ${e}^{v}(\mathbf{X}_i)$ of each single-view has a mean of 0 and a standard deviation of 1, becoming comparable due to the same scale.
\\
\textbf{Special cases in selecting $\lfloor\frac{N}{R}\rfloor$ samples from each cluster.} During the process of selecting $\lfloor\frac{N}{R}\rfloor$ samples from $R$ clusters, we encountered two special cases that need additional designs. We list them below.
\\
\textit{Case 1: $N$ is no smaller than the size of text-similarity-based selected speech-to-text pairs.} We select all text-similarity-based selected speech-to-text pairs and ignore the upper limitation $N$ by skipping MCSS. As a result, all text-similarity-based selected speech-to-text pairs are directly input to CMSN.
\\
\textit{Case 2: $N$ is smaller than the size of text-similarity-based selected speech-to-text pairs, and there exists a cluster with a size smaller than $\lfloor\frac{N}{R}\rfloor$.} We address this case by a greedy-based sample selection algorithm. It greedily selects all samples in a cluster if the cluster size is smaller than a minimum requirement, which is initialized as $r_{min}=\lfloor\frac{N}{R}\rfloor$ and $r_{min}$ is then updated. Finally, the remaining clusters with cluster sizes that are greater than $r_{min}$ will select $r_{min}$ samples from each remaining cluster.
The algorithm is detailed in Algo.~\ref{algo:greedy}.

\begin{algorithm}[!htb]
\caption{Greedy-Based Sample Selection}
\label{algo:greedy}
\begin{algorithmic}[1]
\Require{$R$ clusters with cluster sizes that are $[l_1, l_2, ..., l_R]$ respectively, and a pre-set expected sampling size $N$ that is smaller than the sum of $[l_1, l_2, ..., l_R]$.}
\State Initialize the number of remaining clusters to be selected, $\hat{R}=R$
\State Initialize the number of remaining samples to be selected: $\hat{N}=N$ 
\State Initialize the minimum size requirement for each cluster: $r_{min}=\lfloor\frac{\hat{N}}{\hat{R}}\rfloor$
\State Sort $l=[l_1, l_2, ..., l_R]$ from small to large, and represent their sorted index list as $\hat{l}$, where $l[\hat{l}[i]]\leq l[\hat{l}[i+1]]$
\State Initialize an empty list $p$ to save the cluster index with cluster size smaller than  $r_{min}$
\State Initialize an empty list $r_{sel}$ to save the selected samples
\State Initialize $i=0$
	\While{$l[\hat{l}[i]]<r_{min}$ \& $i\neq R$}
	\State $\hat{l}[i]\rightarrow p$
        \State all samples in $\hat{l}[i]$-th cluster$\rightarrow r_{sel}$
        \State $\hat{N}=\hat{N}-l[\hat{l}[i]]$
        \State $\hat{R}=\hat{R}-1$
        \State $r_{min}=\lfloor\frac{\hat{N}}{\hat{R}}\rfloor$\Comment{Update $r_{min}$}
        \State $i=i+1$
        \EndWhile

\State Initialize $j=0$
        \While{$j\neq R$}
	\If{$\hat{l}[j]$ not in $p$}
        \State $r_{min}$ samples in $\hat{l}[i]$-th cluster$\rightarrow r_{sel}$
        \State $j=j+1$
        \EndIf
        \EndWhile
        
\Ensure $r_{sel}$
\end{algorithmic}
\end{algorithm}

\begin{table*}[tbh]
\small
\centering
\begin{tabularx}{\textwidth}{cXXX}
\hline
\textbf{Dataset} & \textbf{Text Example} & \textbf{Speech Example} & \textbf{Label (Semantics) Example}  \\
\hline
SLURP & event remaining mona Tuesday & a speech respective to the text & \{'scenario': 'calendar'| 'action': 'set'| 'entities': [\{'type': 'event\_name'| 'filler': 'mona'\}| \{'type': 'date'| 'filler': 'tuesday'\}]\} 
\\
\hline
Mini-PS & are there any other comments but you would don't have a any opposition to the language itself it's fine ok ok any other comments ok should we go & a speech respective to the text & N/A 
\\
\hline
SLUE-VoxPopuli & better enforcement of the eu animal welfare legislation is one of the key priorities for animal welfare and the commission has invested substantial resources in pursuit of this aim. & a speech respective to the text & Semantics: \{'entities': [\{'type': ' CARDINAL ’| 'filler': ‘one'\}| \{'type’: GPE'| 'filler': ‘eu'\}]\}
\\
\hline
VoxPopuli  & eu pharmaceutical legislation contains a number of tools to facilitate early access to medicines for patients with unmet medical needs. & a speech respective to the text & N/A
\\
\hline
\end{tabularx}
\caption{Sample examples from each data set used in our experiments.}
\label{table:data_example}
\end{table*}

\subsection{Data Splits and Examples}
\label{app:data_example}
As for the MiniPS2SLURP dataset construction, we sample 40.5\% of SLURP training set for $D^{A \shortrightarrow L, t} $ to train $\Theta^{A \shortrightarrow L, t} $. 
For $D^{A \shortrightarrow T, t}$ and $D^{A \shortrightarrow T, \epsilon}$ used in training $\tilde{\Theta}^{A \shortrightarrow L}$, we use the same 40.5\% of the SLURP training set (having totally same speeches to $D^{A \shortrightarrow L, t}$, but no semantics) and full Mini-PS (32255 pairs) respectively to simulate a real collected speech-to-text pair set $D^{A \shortrightarrow T}$. 

As for the VoxPopuli2SLUE dataset construction, we sample 45\% of SLUE-VoxPopuli fine-tune set for $D^{A \shortrightarrow L, t} $ to train $\Theta^{A \shortrightarrow L, t} $. 
For $D^{A \shortrightarrow T, t}$ and $D^{A \shortrightarrow T, \epsilon}$ used in training $\tilde{\Theta}^{A \shortrightarrow L}$, we use the same 45\% of SLUE-VoxPopuli fine-tune set (having totally same speeches to $D^{A \shortrightarrow L, t}$, but no semantics) and full VoxPopuli (182466 pairs) respectively to simulate a real collected speech-to-text pair set $D^{A \shortrightarrow T}$. 

We list data examples in Tab.~\ref{table:data_example}.

\subsection{License}
\label{sec:license}
Our datasets are  built on the SLUE-VoxPopuli~\cite{shon2022slue} (using CC0 license), VoxPopuli~\cite{wang2021voxpopuli} (using CC BY 4.0 license), SLURP~\cite{bastianelli2020slurp} (using CC BY 4.0 license), and Mini-PS~\cite{galvez2021people} (using CC-BY-SA and CC-BY 4.0 licenses). Considering these licenses, our usage of these existing datasets is consistent with their licenses. According to these licenses, VoxPopuli2SLUE is CC BY 4.0 license, and MiniPS2SLURP is CC-BY-SA and CC-BY 4.0 licenses.

\subsection{Implementation Details}
\label{app:implementation_details}
Our work is implemented on SpeechBrain~\cite{ravanelli2021speechbrain}. The NLU model $\Theta^{T \shortrightarrow L, t}$ is trained by 80\% of $D^{T \shortrightarrow L, t}$ and validated by 10\% of $D^{T \shortrightarrow L, t}$. The SLU model training also uses the same dataset split ratio.
We train NLU for 20 epochs and SLU for 35 epochs, and the parameters performing the best on the validation set will be kept. 
We set the K-Means cluster numbers as 100 in our both two dataset text embedding spaces, where these text clusters will be used for the MCSS as the text modal cluster results of $D^{T\shortrightarrow L, t}$. For MCSS, we set the numbers of audio clusters, semantic types, and multi-view cluster numbers $R$ as 100, 53, 30 in the MiniPS2SLURP setting and 100, 18, and 30 in the VoxPopuli2SLUE, respectively.  
Each of the SLU models and NLU models in our experiments  consists of an encoder and a decoder. Each SLU encoder is the HuBERT encoder~\cite{hsu2021hubert}. Each NLU encoder is either LSTM~\cite{hochreiter1997long} or BERT~\cite{devlin2018bert} encoder. For the SLU and NLU decoders, they are both attentional RNN decoders~\cite{bahdanau2014neural}. To reproduce our main results for both GloVe-based and SentBERT-based in Tab.~\ref{table:main_results}, we set $\beta=\gamma=\alpha=0.1$, $\tau=0.55$, $w^{T}=w^{L}=10$, $w^{A}=1$ and $N=35000$  on MiniPS2SLURP; on VoxPopuli2SLUE, we set $\beta=\gamma=\alpha=0.1$, $\tau=0.75$, $w^{T}=w^{L}=w^{A}=1$ and $N=5556$.

\subsection{Case Study}
\label{app:case_study}
We also show case studies of our $\tilde{\Theta}^{A \shortrightarrow L}$ on the two datasets, shown in Table~\ref{table:case_study_twoset}.

\begin{table*}[!htb]
\small
\centering
\begin{tabularx}{\textwidth}{X|X|X|X|X}
\hline
\textbf{Audio (Shown by its respective text)} & \textbf{Ground-Truth Semantic Label} & \textbf{$\tilde{\Theta}^{A \shortrightarrow L}$ (w/o CMSN, w/o MCSS) Predicted Label} & \textbf{$\tilde{\Theta}^{A \shortrightarrow L}$ (w/o MCSS) Predicted Label}  & \textbf{$\tilde{\Theta}^{A \shortrightarrow L}$ Predicted Label} 
\\
\hline
\hline
\multicolumn{5}{l}{\textit{MiniPS2SLURP}}
\\
\hline
how long does it take to make vegetable lasagna 
& 'scenario': 'cooking', 'action': 'recipe', 'entities': [\{'type': 'food\_type', 'filler': 'vegetable lasagna'\}]
& 'scenario': \textcolor{red}{'news'}, 'action': \textcolor{red}{'query'}, 'entities': [\{'type': \textcolor{red}{'news\_topic'}, 'filler': \textcolor{red}{'election'}\}, \{'type': \textcolor{red}{'date'}, 'filler': \textcolor{red}{'monday'}\}]
& 'scenario': \textcolor{red}{'recommendation'}, 'action': \textcolor{red}{'locations'}, 'entities': [\{'type': \textcolor{red}{'business\_type'}, 'filler': \textcolor{red}{'restaurant'}\}]
& scenario': 'cooking', 'action': 'recipe', 'entities': [\{'type': 'food\_type', 'filler': \textcolor{red}{'cookies'}\}] 
\\
\hline
'remind me the meeting with allen on fifteenth march' 
&'scenario': 'calendar', 'action': 'set', 'entities': [\{'type': 'event\_name', 'filler': 'meeting\}, \{'type': 'person', 'filler': 'allen'\}, \{'type': 'time', 'filler': 'fifteenth march'\}]
& 'scenario': 'calendar', 'action': 'set', 'entities': [\{'type': 'event\_name', 'filler': 'meeting'\}, \{'type': \textcolor{red}{'relation'}, 'filler': \textcolor{red}{'wife'}\}, \{'type': \textcolor{red}{'date'}, 'filler': \textcolor{red}{'march'}\}]
&  'scenario': 'calendar', 'action': 'set', 'entities': [\{'type': 'event\_name', 'filler': 'meeting'\}, \{'type': \textcolor{red}{'date'}, 'filler': \textcolor{red}{'march fifth'}\}]
&  'scenario': 'calendar', 'action': 'set', 'entities': [\{'type': 'event\_name', 'filler': 'meeting'\}, \{'type': 'person', 'filler': 'allen'\}]
\\
\hline
can i please have the weather for tomorrow here in costa mesa
&'scenario': 'weather', 'action': 'query', 'entities': [\{'type': 'date', 'filler': 'tomorrow'\}, \{'type': 'place\_name', 'filler': 'costa mesa'\}]
&'scenario': \textcolor{red}{'calendar'}, 'action': 'query', 'entities': [\{'type': 'date', 'filler': 'tomorrow'\}, \{'type': \textcolor{red}{'time'}, 'filler': \textcolor{red}{'eight am'}\}, \{'type': \textcolor{red}{'date'}, 'filler': \textcolor{red}{'tomorrow'}\}]
&'scenario': 'weather', 'action': 'query', 'entities': [\{'type': 'date', 'filler': 'tomorrow'\}, \{'type': \textcolor{red}{'time'}, 'filler': \textcolor{red}{'nine am'}\}]
&'scenario': 'weather', 'action': 'query', 'entities': [\{'type': 'date', 'filler': 'tomorrow'\}] 
\\
\hline
'should i take my raincoat with me now'
& 'scenario': 'weather', 'action': 'query', 'entities': [\{'type': 'weather\_descriptor', 'filler': 'raincoat'\}]
& 'scenario': \textcolor{red}{'play'}, 'action': \textcolor{red}{'audiobook'}, 'entities': [\{'type': \textcolor{red}{'media\_type'}, 'filler': \textcolor{red}{'audiobook'}\}]
&  'scenario': 'weather', 'action': 'query', 'entities': [\{'type': 'weather\_descriptor', 'filler': \textcolor{red}{'rain'}\}, \{'type': \textcolor{red}{'date'}, 'filler': \textcolor{red}{'today'}\}]
& 'scenario': 'weather', 'action': 'query', 'entities': [\{'type': 'weather\_descriptor', 'filler': \textcolor{red}{'raining'}\}] 
\\
\hline
\hline
\multicolumn{5}{l}{\textit{VoxPopuli2SLUE}}
\\
\hline
second i do not believe in the minsk group but i believe that the eu in the person of the high representative has the capacity to broker the negotiations.
& 'entities': [\{'type': 'gpe', 'filler': 'eu'\}, \{'type': 'org', 'filler': 'minsk group'\}, \{'type': 'ordinal', 'filler': 'second'\}]
& 'entities': [\{'type': 'gpe', 'filler': 'eu'\}, \{'type': 'ordinal', 'filler': \textcolor{red}{'secondly'}\}, \{'type': \textcolor{red}{'ordinal'}, 'filler': \textcolor{red}{'secondly'}\}]
&'entities': [\{'type': 'gpe', 'filler': 'eu'\}, \{'type': 'ordinal', 'filler': \textcolor{red}{'secondly'}\}]
&'entities': [\{'type': 'gpe', 'filler': 'eu'\}, \{'type': 'ordinal', 'filler': 'second'\}]
\\
\hline
what can be done to ensure that the revision process     goes smoothly and is finalised before one may two thousand and fifteen as specified in     article nineteen of the multiannual financial framework regulation so as to avoid losi   ng uncommitted amounts from?
& 'entities': [\{'type': 'law', 'filler': 'article nineteen of the multiannual financial framework'\}, \{'type': 'date', 'filler': 'one may two thousand and fifteen'\}]
& 'entities': [\{'type': 'date', 'filler': \textcolor{red}{'two thousand and twenty'}\}, \{'type': \textcolor{red}{'date'}, 'filler': \textcolor{red}{'two thousand and twenty'}\}]
& 'entities': [\{'type': 'date', 'filler': \textcolor{red}{'two thousand and fifty'}\}
&'entities': [\{'type': 'date', 'filler': \textcolor{red}{'two thousand and fifteen'}\}] 
\\
\hline
\end{tabularx}
\caption{Case studies of $\tilde{\Theta}^{A \shortrightarrow L}$ on two datasets are shown, where red fonts highlight incorrectly predicted tokens. We find that using both MCSS and CMSN (the last column) has the fewest incorrectly predicted tokens. This also verifies the effectiveness of reducing imbalance and noise by our CMSST framework, which includes both MCSS and CMSN.
}
\label{table:case_study_twoset}
\end{table*}

\subsection{Parameter Analysis}
\label{sec:parameters}
\begin{figure*}[!htbp]
\centering
\includegraphics[width=1.0\textwidth]{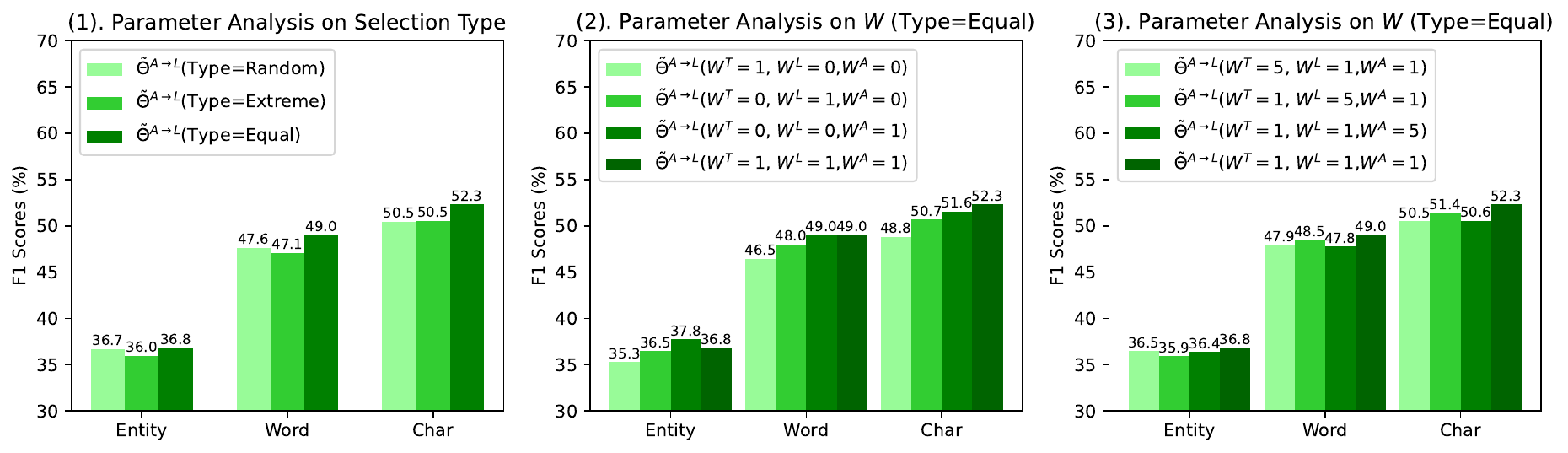}
\caption{Parameter analysis of MCSS on VoxPopuli2SLUE, where BERT-based $\Theta^{T \shortrightarrow L, t}$ is used. All groups have $\|D^{A \shortrightarrow T, t}\|=59$ and $\|D^{A \shortrightarrow T, \epsilon}\|=5461$ for fair comparison.}
\label{fig:paramter_anlysis_mcss_slue}
\end{figure*}

\begin{figure*}[!htbp]
\centering
\includegraphics[width=0.7\textwidth]{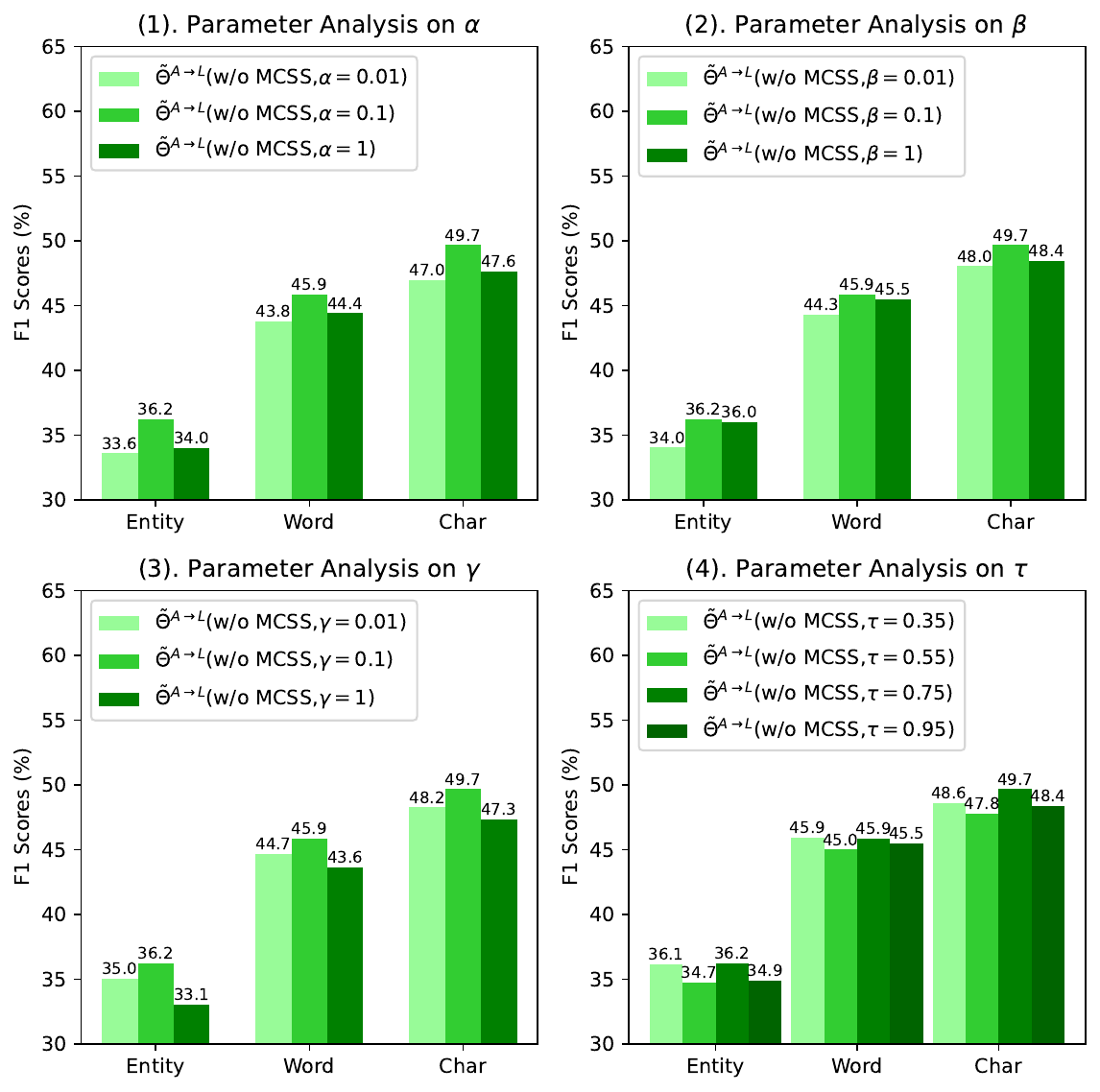}
\caption{Parameter analysis of CMSN on VoxPopuli2SLUE, where LSTM-based $\Theta^{T \shortrightarrow L, t}$ is used. All groups have $\|D^{A \shortrightarrow T, t}\|=68$ and $\|D^{A \shortrightarrow T, \epsilon}\|=5489$ for fair comparison.}
\label{fig:paramter_anlysis_cmsn_slue}
\end{figure*}

The parameter analysis of MCSS and CMSN are respectively shown in Figure~\ref{fig:paramter_anlysis_mcss_slue} and Figure~\ref{fig:paramter_anlysis_cmsn_slue}.

For MCSS, from the Figure~\ref{fig:paramter_anlysis_mcss_slue}, which shows the parameters of the coefficients of MCSS, $w^{T}$, $w^{L}$ and $w^{A}$, we can find below. 
\\
1. $w^{T}$, $w^{L}$, and $w^{A}$ all impact the performance of MCSS. The figure shows performance variant to different weights of $w^{T}$, $w^{L}$, and $w^{A}$. 
\\
2. Considering all three views leads to better performance. Concretely, among the cases shown in the (2) subfigure, we see that $w^{T}=w^{A}=w^{L}=1$ leads to better performance than other single-view cases. This shows the benefit of comprehensively considering three views.

For CMSN, we change one parameter at once and keep the rest of the parameters fixed; we show each of the four parameters on VoxPopuli2SLUE, from which we find that $\beta=\gamma=\alpha=0.1$ and $\tau=0.75$ perform the best.

\subsection{Explanation of Zero-Shot}
\label{sec:why_zero_shot}
Our use of the term “zero-shot” refers to training a spoken language understanding model without ground-truth speech-semantics pairs.

We clarify the reason why there are no ground-truth speech-semantics pairs for SLU, which defines it as a zero-shot setting. Specifically, our SLU model learns from audio-text (A, T) and text-label (T, L) pairs, but lacks audio-label pairs (A, L). Having (X, Z) and (Z, Y) without (X, Y) is a zero-shot problem. For instance, zero-shot evaluations, as seen in Neural Machine Translation (NMT) (e.g.,~\citealp{johnson2017google}), involve training an NMT model with Portuguese→English and English→Spanish examples, which can then generate reasonable translations for Portuguese→Spanish, despite not having seen any data for that language pair. 
Similarly, our work trains an SLU model (e.g., speech→semantics) via speech-text pairs and text-semantics pairs, making it a zero-shot task as no ground-truth speech-semantics pairs are provided.

\end{document}